\documentclass{aa}  
\usepackage{graphicx}
\usepackage{txfonts}

\usepackage{natbib}

\begin{document}

     \title{Non-thermal desorption from interstellar dust grains {\em via} exothermic surface reactions}

     \subtitle{}

     \author{R. T. Garrod\inst{1,2}, V. Wakelam\inst{1,3}
          \and
          E. Herbst\inst{1,4} }
 \offprints{R. T. Garrod, \email{rgarrod@mpifr-bonn.mpg.de}}
   \institute{Department of Physics,
    The Ohio State University, Columbus, OH 43210, USA \and
    Max-Planck-Institut f{\"u}r Radioastronomie, Auf dem H{\"u}gel 69, 53121 Bonn, Germany \and
    Universit\'e Bordeaux 1, L3AB, UMR5804, BP 89 33270 Floirac, France \and
    Departments of Astronomy and Chemistry, The Ohio State University, Columbus, OH 43210,USA}

     \date{Received xxx / Accepted xxx }
     
     \abstract
   {xxx}
   {The  gas-phase abundance of methanol in dark quiescent cores in the interstellar medium cannot be explained by gas-phase chemistry.  In fact, the only possible synthesis of this species appears to be production on the surfaces of dust grains followed by desorption into the gas.  Yet, evaporation is inefficient for heavy molecules such as methanol at the typical temperature of 10 K.  It is necessary then to consider non-thermal mechanisms for desorption.  But, if such mechanisms are considered for the production of methanol, they must be considered for all surface species.}
   {Our gas-grain network of reactions has been altered by the inclusion of a non-thermal desorption mechanism in which the exothermicity of surface addition reactions is utilized to break the bond between the product species and the surface.  Our estimated rate for this process derives from a simple version of classical unimolecular rate theory with a variable parameter only loosely contrained by theoretical work.}
   {Our results show that the chemistry of dark clouds is altered slightly at times up to 10$^{6}$ yr, mainly by the enhancement in the  gas-phase abundances of hydrogen-rich species such as methanol that are formed on grain surfaces.  At later times, however, there is a rather strong change.  Instead of the continuing accretion of most gas-phase species onto dust particles, a steady-state is reached for both gas-phase and grain-surface species, with significant abundances for the former.  Nevertheless, most of the carbon is contained in an undetermined assortment of heavy surface hydrocarbons.}
   {The desorption mechanism discussed here will be better constrained by observational data on pre-stellar cores, where a significant accretion of species such as CO has already occurred.}

     \keywords{Astrochemistry -- ISM: abundances -- ISM: molecules -- molecular processes}

     \titlerunning{Desorption {\em via} exothermic surface reactions}
     \authorrunning{Garrod et al.}

     \maketitle

\section{Introduction}

Methanol (CH$_3$OH) is a molecule commonly detected over a wide range of conditions in interstellar clouds. In quiescent dark cloud regions, it is present in the gas phase with a typical abundance of $\sim$$1.5 \times 10^{-9} n_{\rm H}$ \cite[]{smith04a}. Gas-phase chemical kinetic models have long been successful at reproducing this abundance, invoking the radiative association of CH$_{3}^{+}$ and H$_2$O to form protonated methanol, followed by recombination with electrons to produce methanol and atomic hydrogen. The rate coefficient adopted in both the OSU chemical network ({\em osu.2003}) and the UMIST  rate99 ratefile \cite[]{leteuff00a} for the radiative association reaction was estimated to be $k_{RA} = 5.50 \times 10^{-12} (T/300)^{-1.7}$ cm$^{3}$ s$^{-1}$ (for $T=10 - 50$ K) \citep{bates83a,herbst85a}. However, the rate constant has more recently been experimentally determined by \cite{luca02a}; they obtain an upper limit of $2.0 \times 10^{-12}$ cm$^{3}$ s$^{-1}$ for a temperature of 50 K, with an uncertainty of 30 K. On the assumption that this rate is applicable to temperatures of 10 K, the canonical gas temperature of quiescent dark cloud regions, the measured rate falls short of the estimated value by approximately 3 orders of magnitude. As outlined by \nocite{garrod06b} Garrod et al. (2006a, hereafter GPCH) such rates are incapable of reproducing observed dark cloud methanol abundances.

In addition, the rate coefficient and branching ratios of the dissociative recombination of CH$_{3}$OH$_{2}^{+}$ and CD$_{3}$OD$_{2}^{+}$ were recently measured by \cite{geppert06a}. Whilst their experiment suggested a rate constant at 10 K a few times larger than that previously adopted in the reaction networks, they measured a branching fraction for CH$_3$OH of $3 \pm 2$ \%
($6 \pm 2$ \% 
for CD$_3$OD), as compared to an assumed 50 -- 100 \% 
in chemical models.
The strongest channels produce multiple fragments, mainly consisting of separate carbon- and oxygen-groups.
The combination of these new rates and branching ratios leaves the gas-phase route incapable of producing the observed methanol abundances; peak model values are at least five orders of magnitude short of the target. Even with an alternative to the radiative association -- a reaction between formaldehyde and protonated formaldehyde to produce protonated methanol and carbon monoxide -- the modelled values are entirely inadequate.

Hence, in the absence of some heretofore unexplored gas phase reaction, we must conclude that grain-surface formation of methanol is responsible for gas phase abundances. Surface methanol is typically assumed to be formed {\em via} the repetitive hydrogenation of accreted CO by atomic hydrogen, which is comparatively mobile on dust-grain surfaces, even at 10 K \cite[see][]{katz99a}. Experimental work by \cite{watanabe02a} and \cite{hidaka04a} has demonstrated that this process is efficient on ice surfaces at such low temperatures, in spite of activation energies on the order of 1000 K for the reactions H + CO $\rightarrow$ HCO and H + H$_2$CO $\rightarrow$ CH$_3$O/CH$_2$OH. Methanol is indeed detected in icy dust-grain mantles in infrared absorption towards background stars. \cite{nummelin01a} suggest an abundance of $< 4 \times 10^{-6}$ $n_H$ along the line of sight to the background star Elias 16. The OSU gas-grain chemical code \cite[]{hasegawa92a}, which uses rate equations to model the coupled gas-phase and grain-surface chemistry, is capable of producing surface-bound methanol in such quantities, hence we would require only around 0.1\% 
of total methanol formed in this way to be present in the gas-phase to reproduce observed gas-phase abundances. But, crucially, thermal desorption at 10 K is negligible for species more massive than molecular hydrogen; we therefore require some other means of returning surface-produced methanol to the gas phase.

\begin{table}
\caption[]{Models}
\label{tab2}
\begin{center}
\begin{tabular}{ll}
\hline
\hline
\noalign{\smallskip}
Model  & $a$ \\
\noalign{\smallskip}
\hline
\noalign{\smallskip}
M0 & 0 \\
\noalign{\smallskip}
M1 & 0.01 \\
\noalign{\smallskip}
M2 & 0.03 \\
\noalign{\smallskip}
M3 & 0.1 \\
\noalign{\smallskip}
\hline
\end{tabular}
\end{center}
\end{table}

 Certain non-thermal desorption mechanisms have been implemented in chemical models; however, only cosmic ray-induced heating desorption is frequently employed in dark cloud models \cite[]{hasegawa93a}, although even this mechanism appears too slow to produce appreciable methanol desorption. Some models utilise a direct photo-desorption mechanism \cite[]{draine79a,hartquist90a,willacy98a}, but the yield per photon, and therefore the overall rate, are uncertain. 

Some work has been conducted on other possible mechanisms. \cite{williams68a} first suggested that the stabilisation of the excited product of an exothermic grain surface reaction could result in its desorption from the surface. This idea was considered further by \cite{watson72a,watson72b}, but no firm desorption fraction was suggested; however, their estimates were more efficient than we require in this model. \cite{jones84a} calculated that the products of the reactions O + H $\rightarrow$ OH and OH + H $\rightarrow$ H$_2$O should be retained on the grains with a probability of between 70 and 100\% for grain-mantle formation to be efficient in dark clouds. 

\cite{duley93a} suggested that the highly exothermic reaction between hydrogen atoms on grain surfaces, forming ro-vibrationally excited H$_2$,
may impart some energy to the grain surface. This would allow the localised evaporation of weakly bound molecules like CO in regions where H$_2$ formation had not yet reduced gas-phase atomic hydrogen abundances to their low steady-state levels. 

\begin{table}
\caption[]{Initial abundances of H$_2$ and elements \cite[]{graedel82a}.}
\label{tab1}
\centering
\begin{tabular}{l c}
\hline\hline
\noalign{\smallskip}
Species $i$ & $n_{i}/n_{H}$ $^{\dag}$ \\
\noalign{\smallskip}
\hline
\noalign{\smallskip} 
H$_2$ & $0.5$ \\
\noalign{\smallskip}
He & $0.14$ \\
\noalign{\smallskip} 
C$^+$ & $7.3(-5)$ \\
\noalign{\smallskip}
N & $2.14(-5)$ \\
\noalign{\smallskip}
O & $1.76(-4)$ \\
\noalign{\smallskip}
S$^+$ & $8.0(-8)$ \\
\noalign{\smallskip}
Na$^+$ & $2.0(-9)$ \\
\noalign{\smallskip}
Mg$^+$ & $7.0(-9)$ \\
\noalign{\smallskip}
Si$^+$ & $8.0(-9)$ \\
\noalign{\smallskip}
P$^+$ & $3.0(-9)$ \\
\noalign{\smallskip}
Cl$^+$ & $4.0(-9)$ \\
\noalign{\smallskip}
Fe$^+$ & $3.0(-9)$ \\
\noalign{\smallskip}
\hline
\noalign{\smallskip}
$^{\dag}$ $a(b) = a \times 10^{b}$ \\
\end{tabular}
\end{table}

In this study, we further investigate the non-thermal desorption mechanism presented by GPCH, in which the energy of formation released by the association reaction of grain-surface radicals may break the surface--molecule bond of the product. The probability for this to occur is small (on the order of 1\%), 
and therefore most of the product remains on the grain surface. The molecular dynamics study of \cite{kroes06a} also suggests that this mechanism may be important for photodissociated water ice. Here we apply the mechanism not only to methanol production but to all surface reactions which result in a single product.  A statistical theory is used to determine the probability of desorption, which is dependent on the product species and the exothermicity of the reaction. 

We use the OSU gas-grain chemical code to investigate the detailed chemistry which takes place both in the gas-phase and on the grain-surfaces as a result of the new mechanism.
In addition, by comparison with observed abundances, we weakly constrain the generalised free-parameter, $a$, which strongly determines the desorption probability for all qualifying reactions.

\begin{figure*}
\centering
\includegraphics[width=8.5cm]{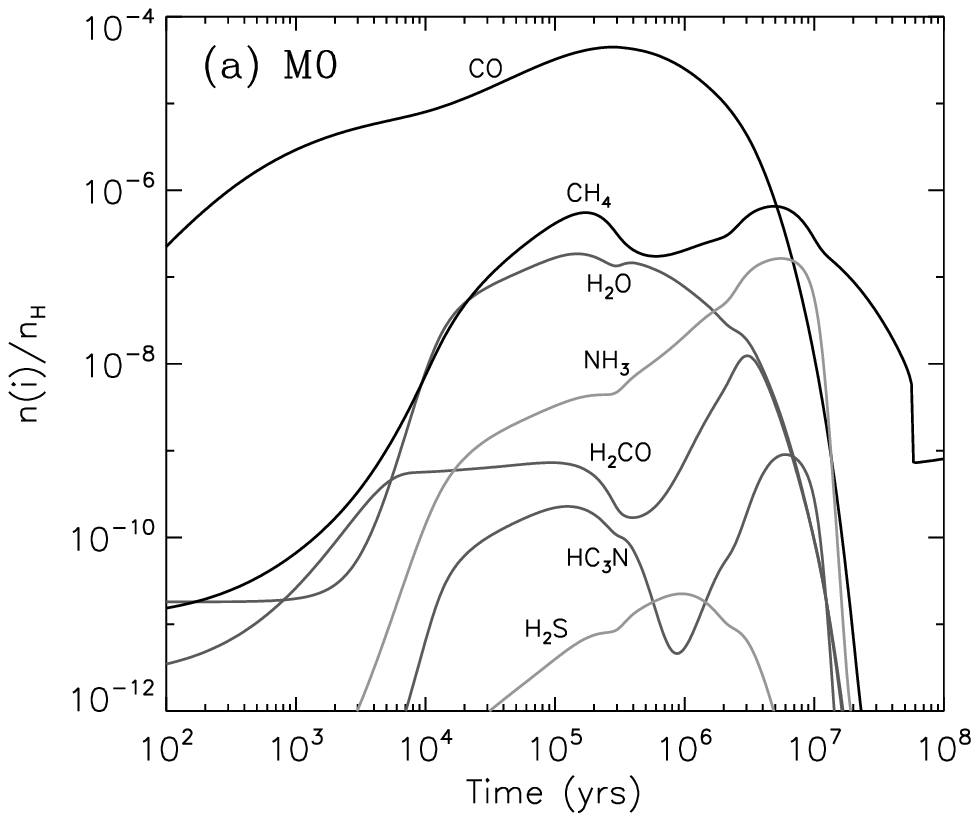}
\includegraphics[width=8.5cm]{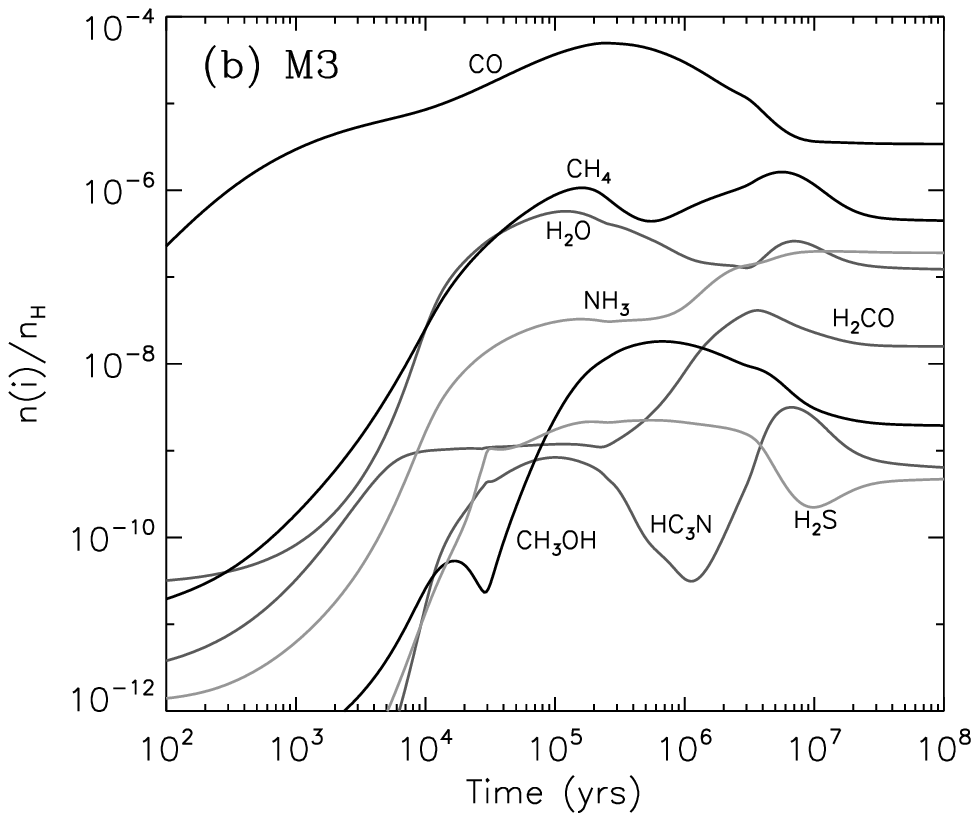}
\caption{Gas phase abundances for models (a) without the new desorption mechanism, M0, and (b) with the desorption mechanism at its strongest, M3.}
\label{fig1}
\end{figure*}

In Section 2 we describe the chemical model and the new desorption mechanism. We present the results in Section 3, and compare with observations of L134N and TMC-1CP in Section 4. We present our conclusions in section 5.

\section{Chemical model}

\subsection{The gas-grain code}

To model the chemistry of a dark cloud, we model the gas-phase and grain-surface chemistry in tandem using rate equations. We allow accretion onto the grain surfaces from the gas phase. Surface-bound species may evaporate thermally, according to a Boltzmann law, or by the sporadic heating of the grains by cosmic-ray impacts \cite[]{hasegawa93a}, which is averaged over time. Surface species may be photodissociated, by both the cosmic ray-induced radiation field of \cite{prasad83a} and by the (heavily extinguished) external radiation field. The surface photodissociation rates are identical to their gas-phase analogues, as explained by \cite{ruffle01a}.

Surface species may react with each other; rates are dependent on the concentrations, and the sum of the diffusion rates of the reactants. As explained by \cite{hasegawa92a}, the diffusion rate is defined as the frequency at which the species may thermally hop over the barrier between sites, divided by the total number of sites on the grain surface.

\subsection{Desorption {\em via} exothermic surface reactions}

In addition to the two basic desorption mechanisms, we include that introduced by GPCH. For each surface reaction that leads to a single product, we apply a branching ratio that this product may be ejected into the gas phase, on the assumption that there exists a probability that the energy released in the formation of the molecule acts to break the surface--molecule bond. We assume, in the case of two-product reactions, that the energy is lost to lateral translation along the grain surface; hence, in this case, no desorption occurs.

To quantify the probability of desorption, we apply Rice-Ramsperger-Kessel (RRK) theory \cite[see e.g.][]{holbrook96a}. Modelling the surface--molecule bond as an additional molecular vibrational mode, RRK gives the probability, $P$, for an energy $E > E_D$ to be present in the bond, from a total energy $E_{reac}$:
\begin{equation}
P = \left[  1 - \frac{E_{D}}{E_{reac}}  \right]^{s - 1}
\end{equation}
\noindent where $E_D$ is the desorption energy (binding energy) of the product molecule, $E_{reac}$ is the energy of formation released in the reaction, and $s$ is the number of vibrational modes in the molecule/surface-bond system. This number is $s = 2$ for diatomic species; for all others, $s = 3 N - 5$, where $N$ is the number of atoms in the molecule, which is assumed to be non-linear.  In the hypothetical case where the molecule has no other means of energy loss (i.e. to the grain surface), we would expect desorption at a rate $\nu P$, where $\nu$ is the frequency of the surface--molecule bond (comparable to a Van der Waals bond).

\begin{figure*}
\centering
\includegraphics[width=8.5cm]{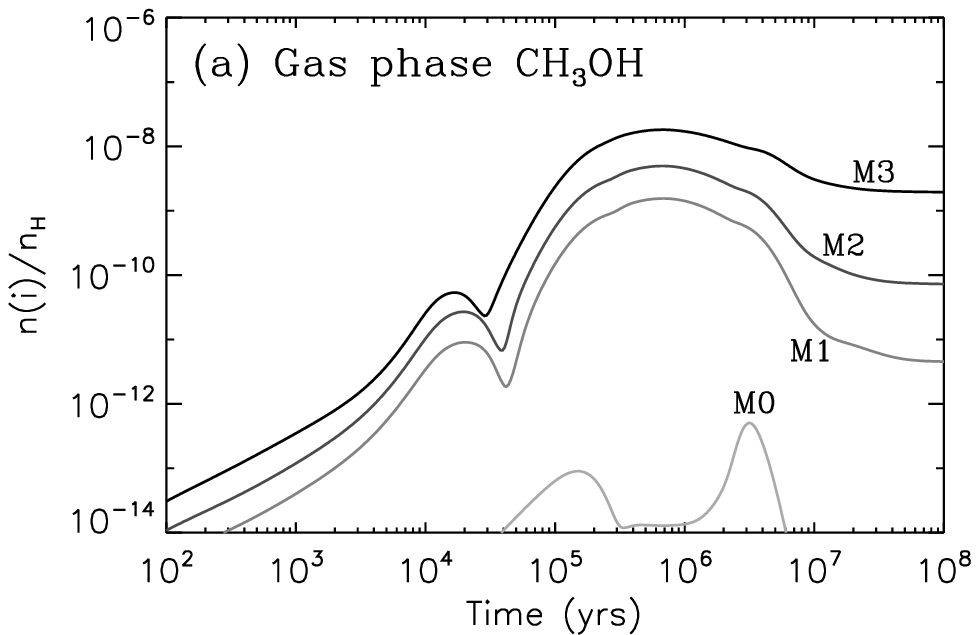}
\includegraphics[width=8.5cm]{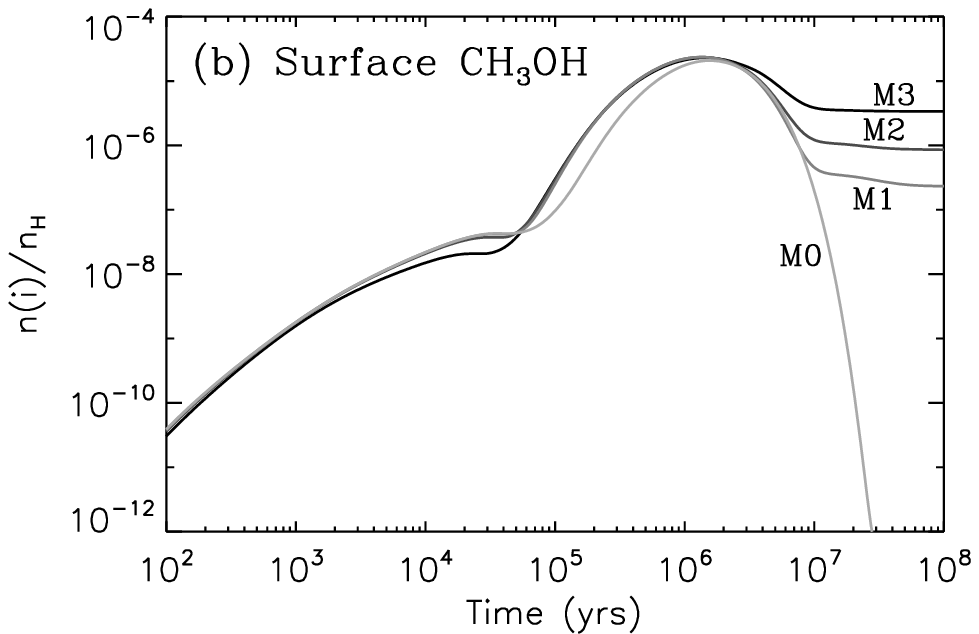}

\caption{CH$_3$OH abundances.}
\label{fig2}
\end{figure*}

\begin{figure*}
\centering
\includegraphics[width=8.5cm]{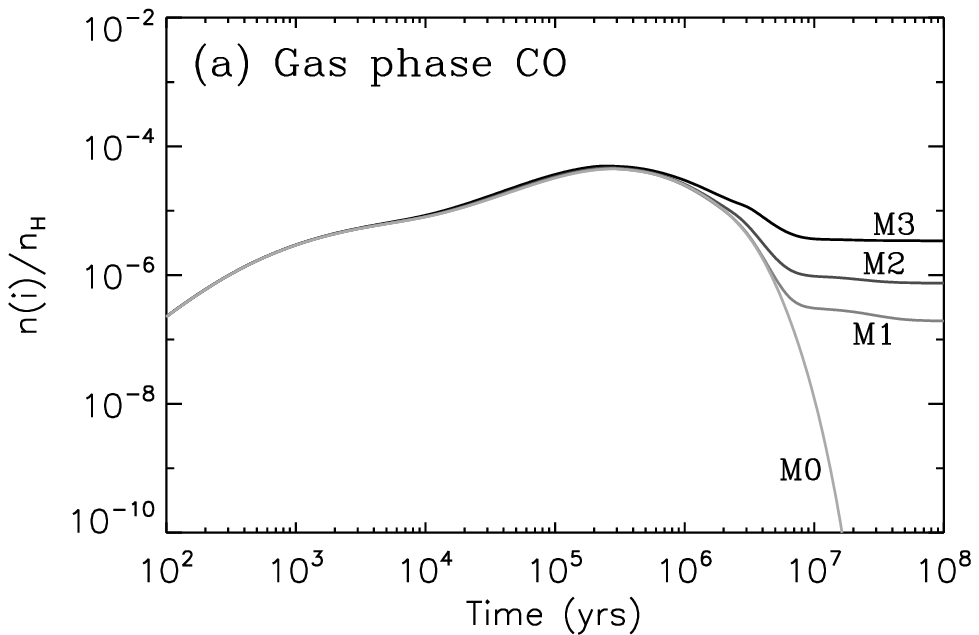}
\includegraphics[width=8.5cm]{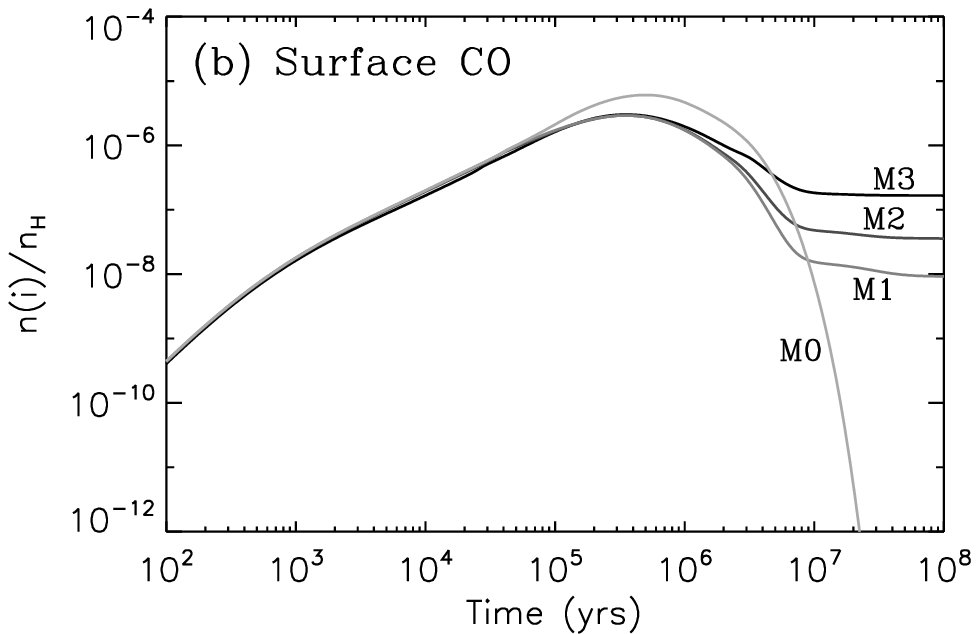}
\caption{CO abundances.}
\label{fig3}
\end{figure*}

In practice, we expect energy loss to the surface to be fast, so in order to obtain the branching fraction for desorption, we must model the competition between the two processes. We define a rate at which the total energy is lost to the surface, $\nu_s$, hence the fraction of reactions resulting in desorption is:
\begin{equation}
f = \frac{\nu P}{\nu_{s} + \nu P} = \frac{a P}{1 + a P}
\end{equation}
\noindent where $a=\nu / \nu_{s}$, the ratio of the surface--molecule bond-frequency to the frequency at which energy is lost to the grain surface. (Whilst the use of a value $a \neq 1$ strictly constitutes an empirical modification to the pure RRK treatment, such modifications are frequently employed to ensure agreement with experiment in other applications of the theory; see \cite{allain96a}, and references therein, for a discussion of this issue). In this study we use a generic value of $a$ for all product species.  Since $E_{D}$ is normally much less than $E_{reac}$, $P$ is approximately unity and $f \approx a$, for small $a$.

The new mechanism is incorporated into the code such that a fraction, $f$, of the product species in qualifying reactions is desorbed, whilst the rest, ($1 - f$), remains as a surface-bound product.

GPCH assume $a=0.1$, but this value has been labelled high \citep{pilling06}.  Also, \cite{kroes06a} have conducted molecular dynamics simulations of the irradiation of water ice with UV photons. They measure the occurrence of a number of possible outcomes, including that in which the H and OH resulting from photodissociation of H$_2$O recombine. From their data, we estimate that $\sim$0.9 \% 
of recombinations result in desorption. Using our value of $E_D$(H$_2$O$) = 5700$ K and $E_{reac} = 5.91 \times 10^4$ K, this implies $a=0.012$. 
In order to test the effects of the new mechanism, and constrain the value of $a$, we investigate models (see Table \ref{tab2}) with various values of $a$: 0, 0.01, 0.03 and 0.1.

\subsection{Rates and initial conditions}

 For this study, we adopt our latest gas-phase chemical network, {\em osu.2005}. The full gas phase ratefile and documentation of updates are available  at {\em http://www.physics.ohio-state.edu/$\sim$eric/research.html}. We have updated the grain surface photodissociation rates in line with the gas-phase values included in the new network.
To calculate the various desorption rates, and the diffusive reaction rates, we use the binding energies, $E_{D}(i)$, and diffusion barriers, $E_{B}(i)$, adopted by \cite{garrod06c}, corresponding to an amorphous water ice surface. These values are typically a little larger than the bare-grain values used by GPCH and earlier models (Cuppen \& Herbst, in prep.). The review of experimental evidence by \cite{katz99a} indicates that quantum tunnelling through diffusion barriers is inefficient, even for atomic hydrogen, therefore all diffusive rates are based on a thermal hopping rate. We treat the so-called modified rates \cite[]{caselli98a,stantcheva01a}, including reactions with activation energy barriers, in the same way as \cite{garrod06c}; all surface atomic hydrogen reaction rates may be modified. At 10 K, surface reactions are dominated by H-addition; other species are much less mobile.
As per \cite{garrod06c}, we assume an activation energy, $E_{A} = 2500$ K \citep{woon02,ruffle00a}, for both of the reactions H + CO $\rightarrow$ HCO and H + H$_2$CO $\rightarrow$ CH$_3$O.

Initial abundances correspond to the so-called low metal abundances of \cite{graedel82a}; see table \ref{tab1}. Physical conditions remain constant throughout the model, with a density $n_H = 2 \times 10^{4}$ cm$^{-3}$, gas and grain temperatures $T_K = T_g = 10$ K, and a visual extinction $A_V = 10$. The cosmic-ray ionisation rate is set to the canonical value, $\zeta = 1.3 \times 10^{-17}$ s$^{-1}$.

\begin{figure}
\centering
\includegraphics[width=8.5cm]{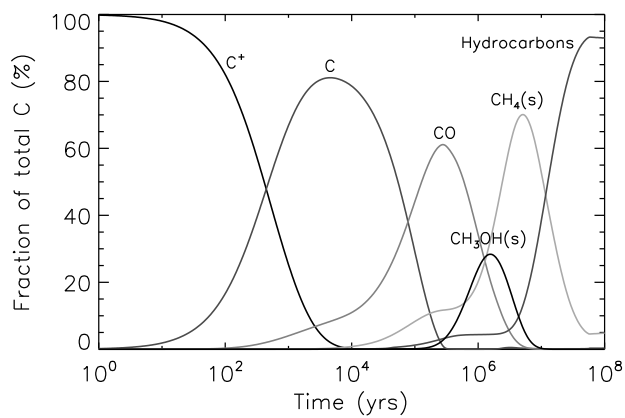}
\caption{Fraction of carbon contained by most important carbon-bearing species for model M0. Surface-bound species are represented by (s).}
\label{fig4}
\end{figure}

\section{Results}

Figures \ref{fig1}a \& b show the calculated abundances of a selection of gas phase species for the two extremes of the model: without the new desorption mechanism, model M0, and with the new desorption mechanism at its strongest, i.e. $a = 0.1$, model M3. These plots show the results of using our amorphous water ice surface, rather than the silicaceous/carbonaceous surface used in previous dark cloud gas-grain models (GPCH). Most species are more tightly bound to the surface, resulting in smaller late-time abundances. Indeed, in model M0, where the strongest desorption mechanism is that caused by cosmic ray-induced heating, desorption is now insignificant for all but the lightest species. For example, the new CR-induced desorption rate for methanol is negligible, using the measured binding energy of 5530 K \cite[]{collings04a}. 

The effects of the new mechanism at late times are quite stark  -- model M3 does not display the total gas phase depletion of heavy species seen in previous models, but reaches an approximate steady state by a time $t = 10^8$ yr. Models M1 ($a = 0.01$) and M2 ($a = 0.03$) also reach steady state, but abundances are proportionately lower, by approximately 10$\times$ and 3$\times$, respectively. Thus, for the majority of gas phase species, the strength of the effect of the new mechanism is quite predictable at very late times. 

At times before $\sim$1 Myr, the majority of species do not exhibit very significant variations between models, although many species do show a somewhat greater peak value (at around 10$^6$ -- 10$^7$ yr) for $a \neq 0$. Since, at the grain temperature of $T_{g} = 10$ K, hydrogenation is the dominant chemical reaction, the effect of the new desorption mechanism is most strongly represented in the gas phase by hydrogen-bearing species.   Most strongly affected are the multiply-hydrogenated species, since their partially hydrogenated precursors are also ejected into the gas phase, where their hydrogenation may also be completed. The peak H$_2$O abundance is raised due to the formation and partial desorption of large amounts of OH and water from the grain surfaces; however, water is still the dominant ice component, and the build-up of water ice mantles is not significantly affected at any time by the new mechanism. CO is also slightly enhanced at its peak, partially due to the desorption of OH, which facilitates reaction with ionised carbon. Naturally, the desorption of OH tends to increase the abundance of O$_2$ (not shown), {\em via} reaction with atomic oxygen. H$_2$S is very strongly enhanced, and is a special case, as discussed in section 3.2.

\begin{figure*}
\centering
\includegraphics[width=13.0cm]{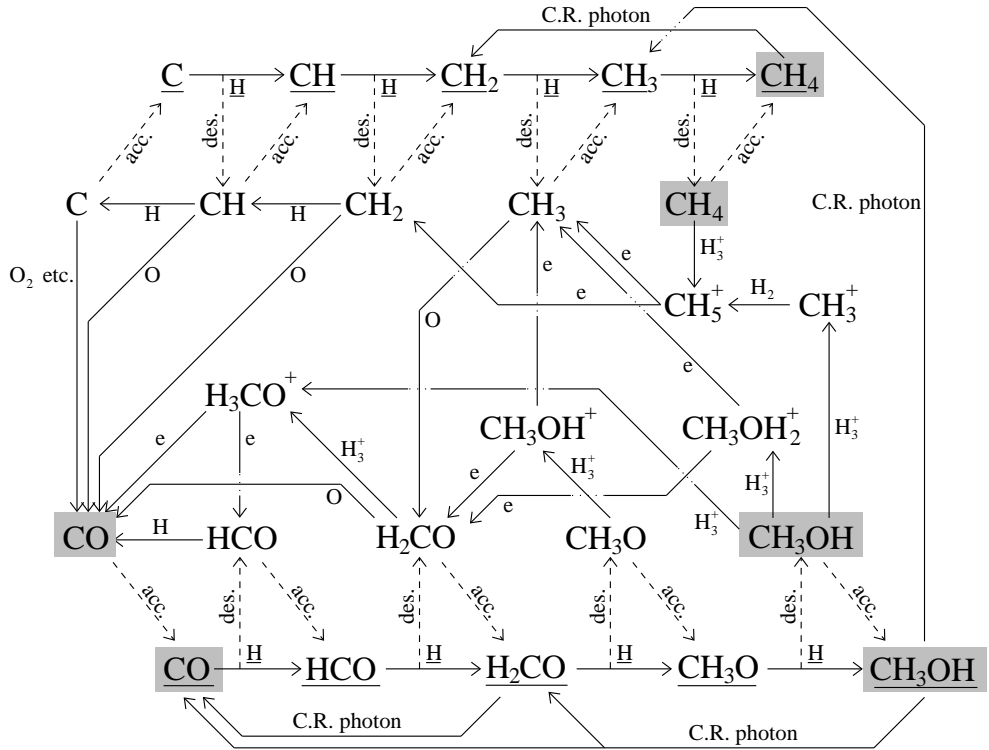}
\caption{Chemical routes involving the conversion among solid and gaseous methanol, carbon monoxide, and methane.  Grain-surface species are underlined.}
\label{fig5}
\end{figure*}

\subsection{Selected carbon- and oxygen-bearing species}

 The peak methanol abundance produced in GPCH with parameter $a = 0.1$ is somewhat higher than observations suggest, typically a few $\times 10^{-9} n_{\rm H}$.  In model M3, this effect is mitigated to some degree by the new stronger binding energies; the resultant peak abundance, $1.5 \times 10^{-8} n_{\rm H}$, is high, but just within an order of magnitude of the typical levels. 
Figure \ref{fig2} shows gas-phase and grain-surface methanol abundances for each model. With $a \neq 0$, gas-phase methanol abundances scale approximately with $a$; model M1 achieves $1.8 \times 10^{-9} n_{\rm H}$ at peak. The range of peak values for models M1 -- M3 confirms the success of the new desorption mechanism in producing appropriate levels of gas-phase CH$_3$OH.

The surface methanol abundance is largely unaffected by the new mechanism until very late times. Interestingly, at these late times the surface CH$_3$OH abundance increases with increasing $a$, so that a larger desorption parameter does not result in lower surface methanol levels; indeed they are strongly enhanced. As can be seen from Fig. \ref{fig3}, the behaviour of CO, the other main repository of grain-surface carbon, is very similar to that of methanol. Figures \ref{fig2} and \ref{fig3} also show commensurate increases in gas phase CO and CH$_3$OH. {\em What is happening  at late times?} 

To answer this question, it is first necessary to consider the major carbon-bearing species as a function of time. Figure \ref{fig4} shows  these dominant  species, for model M0 ($a=0$). We see that in this model (and in models M1 -- M3), most of the carbon does not end up in CO or CH$_{3}$OH but at late times goes through a period in which solid methane dominates before ending up in the form of an assortment of larger surface-bound hydrocarbon molecules.  The state of hydrogenation of such surface-bound hydrocarbons is poorly constrained in our model. The long timescales required ($t > \sim$$10^7$ yr) may of course make the attainment of large hydrocarbon abundances impossible, due to dynamical considerations.

The conversion of surface CO and CH$_{3}$OH into methane {\em via} a variety of processes is depicted in Fig. \ref{fig5}. The methane itself is largely formed by hydrogenation of the methyl (CH$_3$) surface radical. Without the new desorptive mechanism, cosmic-ray photodissociation of CH$_3$OH is the primary formation route for CH$_3$ at later times; the methanol CR-photodissociation branches that produce CO and H$_2$CO quickly result in re-hydrogenation, producing no net effect. Cosmic-ray photodissociation of methane also quickly leads to its own re-formation. 

With the new mechanism activated, the conversion of CO and CH$_3$OH to CH$_4$ occurs during continual recycling between gas and grain surface -- the direct dissociation of methanol into CH$_3$ is no longer dominant. These routes rely heavily on the initial dissociation of surface methanol by cosmic-ray induced photons and desorption {\em via} the new mechanism, leading to a variety of gas-phase species that undergo an assortment of ion-molecule and neutral-neutral processes or re-accrete onto grains. The gas-phase processes eventually produce CH$_{3}$ among other species. This radical then can accrete onto the surface as a precursor for surface methane and, {\em via} the new desorptive mechanism, gas-phase methane.  Alternatively, it can react with atomic C to form acetylene (C$_{2}$H$_{2}$) and H atoms in the gas, leading to more complex hydrocarbons.  In the gas phase, methane can also be converted into larger hydrocarbons by a variety of processes. 

Whilst without the new mechanism, most surface methanol is ultimately channelled into CH$_4$ at late times, the new mechanism allows carbon hydrides to return to the gas phase, where they may be converted into CO or its hydrides, maintaining modest levels of CO, formaldehyde and methanol in both the gas phase and on grain surfaces. The continuous CR-photodissociation of surface CH$_4$ (and other hydrocarbons) keeps this cycle active. These processes are also shown in Fig.~\ref{fig5}.

We may compare the late-time behaviour of our new model with the results of \cite{willacy94a}. They investigated the mechanism of \cite{duley93a}, whereby the formation of H$_2$ releases energy into the grain surface, allowing localised evaporation of weakly bound molecules, most notably CO. Whilst we use a much larger gas-phase and grain-surface reaction network, whose rates have also evolved somewhat in the intervening period, we may comment broadly on the most obvious differences. Whereas we find that complete freeze-out onto grains is prevented by our new mechanism, theirs has the effect of merely retarding freeze-out. This is due to our inclusion of grain-surface photo-destruction processes, which allows re-hydrogenation of the products and the associated evaporation {\em via} the new mechanism. The mechanism of Willacy et al. allows significant re-injection of mantles into the gas phase only while the accreted species are small, with low binding energies, such that the formation energy of reaction H + H $\rightarrow$ H$_2$ may desorb them. When CO is converted in the gas phase into OH and H$_2$O, and CH$_4$ and C$_2$H$_2$, and subsequently accretes, the mechanism is no longer effective and freeze-out takes over. Note also that surface CO is left unprocessed in their model, which tends to prolong the efficiency of their mechanism in returning carbon and oxygen to the gas phase.

\begin{figure*}
\centering
\includegraphics[width=8.5cm]{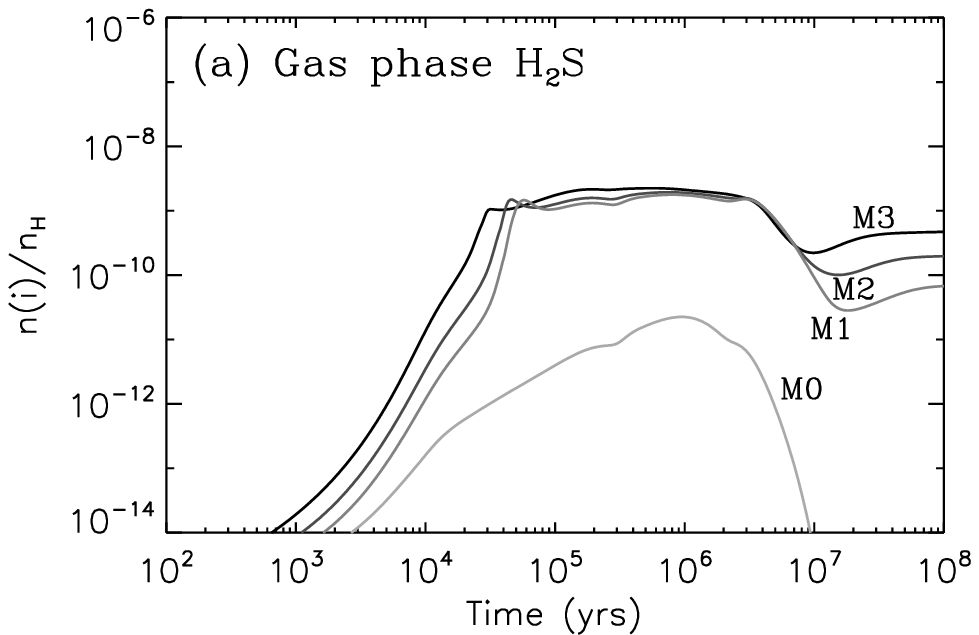}
\includegraphics[width=8.5cm]{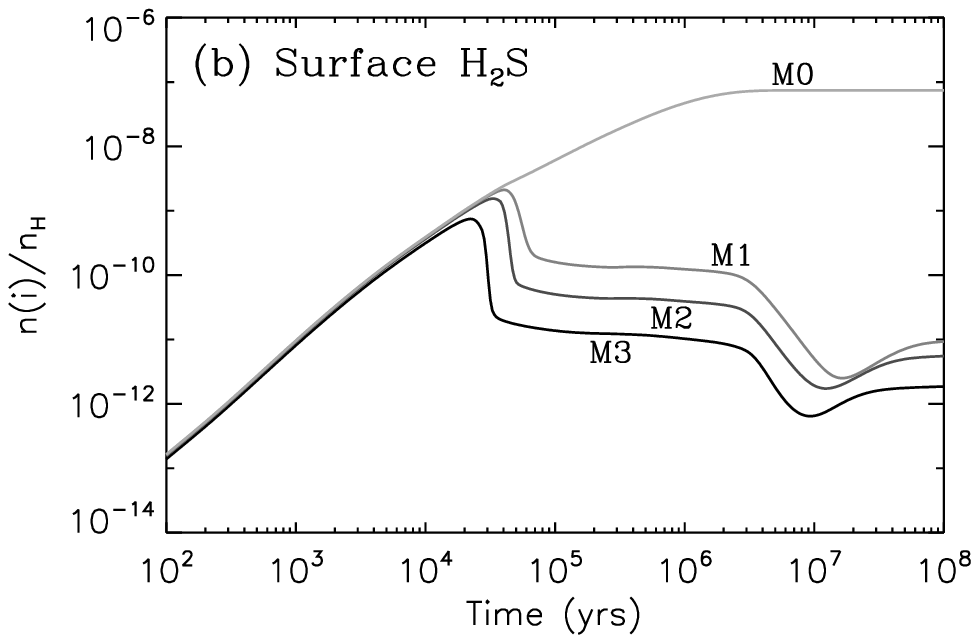}
\caption{H$_2$S abundances.}
\label{fig6}
\end{figure*}

\subsection{Sulphur-bearing species}

The behaviour of sulphur is a long-running problem in interstellar chemistry.  For example, gas-phase SO and CS are typically over-produced.  
Previous gas-grain models have predicted large quantities of H$_2$S on grain surfaces; however, no solid phase H$_2$S has been detected to date in the interstellar medium.  As in this model, the initial elemental sulphur abundance is therefore usually reduced by around 2 orders of magnitude on the diffuse cloud abundance. 

Figures \ref{fig6}a \& b show the gas-phase and grain-surface abundances of H$_2$S for each model. Model M0 produces too much surface H$_2$S (i.e. a significant proportion with respect to H$_2$O ice), and too little in the gas phase. However, the introduction of the new mechanism drastically reduces the abundance on the surface, and greatly increases the gas-phase level, now much more in line with typical observed values \cite[e.g. $8 \times 10^{-10} n($H$_{2})$ in L134N,][]{ohishi92a}, from around $3 \times 10^{4}$ yr onwards. Surface abundances are around $10^{-5}$ of the H$_2$O ice abundance, even for model M1, acceptably low to agree with observations.  The inclusion of the new mechanism, for the range $a = 0.01$ -- $0.1$, also produces abundances of gas-phase CS and SO  in good agreement with observational values at appropriate times, albeit using a much depleted initial abundance of atomic sulphur.

The unusual strength of the effect on H$_2$S is due to the hydrogen abstraction reaction H + H$_2$S $\rightarrow$ HS + H$_2$. H$_2$S is formed by repetitive hydrogenation of atomic sulphur on the grain surfaces. Because the abstraction reaction involves atomic hydrogen, and has only a fairly small activation energy (860 K), it is relatively fast. In model M0, the ratio of HS to H$_2$S achieves a quasi-equilibrium, according to the relative hydrogenation and abstraction rates, resulting in the steady formation of H$_2$S in tandem with H$_2$O. For models M1 -- M3, although the abstraction reaction itself does not (in our model) directly lead to desorption {\em via} the new mechanism, the resultant HS is quickly hydrogenated and this reaction does lead to desorption. Sulphur, in its atomic and hydrogenated forms, is stuck in this fast loop, and so the new mechanism is very efficient at ``syphoning off'' sulphur back into the gas phase. The result is that at times from around $10^{4}$ -- $3 \times 10^{6}$ yr, gas phase atomic sulphur becomes the dominant sulphur-bearing species as the desorbed H$_2$S is broken down. Through the rapid H addition/abstraction rates, the process of removal is efficient enough to ensure a minimal $a$-dependence for gas-phase H$_2$S abundances; no significant build-up of grain-surface sulphur occurs for models M1 -- M3. In all models, surface-bound H$_2$CS becomes the dominant form of sulphur, at times later than $\sim$5 Myr.

The process of H$_2$S formation and re-formation on grains may have implications for the degree of deuteration detected in this species. The speed of the hydrogenation and abstraction reactions would allow a quasi-equilibrium between deuterated and undeuterated forms to arise quickly; deuterium fractionation should therefore be highly sensitive to the branching ratios of the abstraction reactions (e.g. for H + HDS, the ratio of products H$_2$ + DS, versus HD + HS), as well as the specific activation energy barriers for each reaction.

\subsection{Nitrogen-bearing species}

The effects of the new mechanism on nitrogen surface chemistry are not very substantial -- the majority of nitrogen still ends up as NH$_3$ on the grains, a result in some disagreement with observation \citep{whittet03a}. When ammonia is broken down on grain surfaces by cosmic ray-induced photodissociation into NH$_2$ or NH, these products can be hydrogenated, to ammonia and NH$_2$ respectively, and can be desorbed according to the new mechanism. However, the gas-phase chemistry also tends to favour the formation of ammonia from NH$_{2}$, and the ammonia may ultimately re-accrete. Even for the $a = 0.1$ case, the primary form of surface nitrogen is unchanged.

\begin{figure}
\centering
\includegraphics[width=8.5cm]{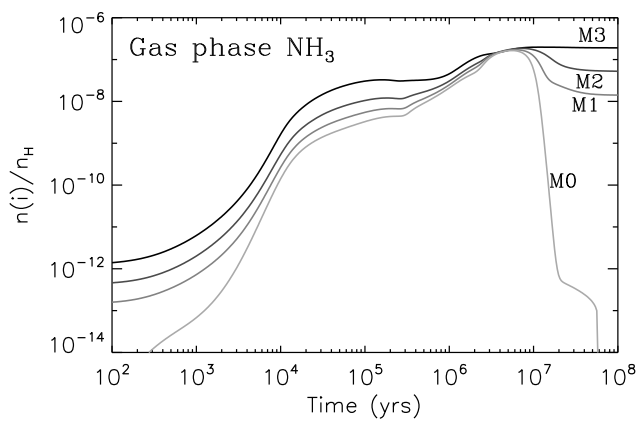}
\caption{NH$_3$ abundances.}
\label{fig7}
\end{figure}
An important effect is the raised abundance of gas-phase NH$_3$  at early times (see Fig.~\ref{fig7}), due to the very moderate fractions released from the grain surfaces. At nearer to 1 Myr, the difference among models tails off, and the peak value, at $t \simeq 5$ Myr, is unaffected. Clearly at times of 10$^7$ yr or greater, there is a very large disparity between M0 and the models which include the new mechanism. 
Hence, the new desorption mechanism has the effect of making NH$_3$ more of an ``early-time'' species, although the effect is very significant only for the largest $a$-value.

The cyanopolyynes included in the code (HC$_3$N, HC$_5$N, HC$_7$N \& HC$_9$N) are all enhanced  at both their early-time peak and late-time peak; the enhancement is an order of magnitude for model M3. This effect arises primarily from the injection of hydrocarbons into the gas phase, allowing increased reaction with atomic nitrogen.

\section{Comparison with observations in quiescent dense cores}

The analysis of GPCH included a comparison of the model with the observational results for TMC-1CP. This was done in the same fashion as \cite{smith04a}, simply by comparing the number of species with computed abundances that fell within one order of magnitude of the detected level. This method is helpful in comparing models which are very different, i.e. gas phase vs. gas-grain, however it is less useful for the comparison of the range of models which we exhibit here, which are mostly quite similar. It is also harder to distinguish the {\em time} of best fit for a particular model. The problem arises from setting such a strict criterion for a ``success''; inevitably, some species fall close to the order of magnitude limit, and so similar models can produce wildly different levels of ``success''. We therefore require a quantitative means of evaluation of the success of the model in reproducing observations.

\begin{figure}
\centering
\includegraphics[width=8.5cm]{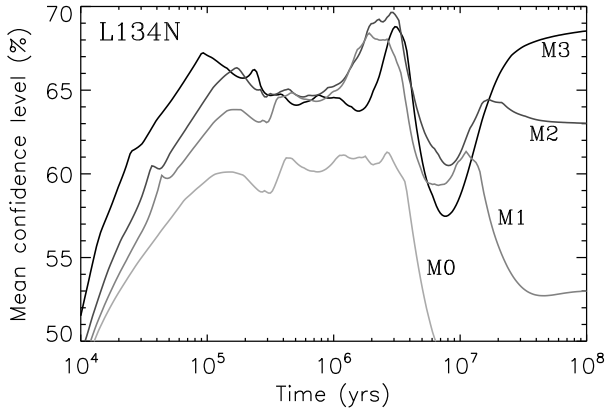}
\caption{Mean confidence level (\%) for fit with L134N observations.}
\label{fig98}
\end{figure}

As a different approach, \cite{wakelam06b} compared observations with the results of a gas-phase model taking into account both observational and theoretical error bars, the latter computed using statistical methods based on uncertainties in rate coefficients.  Agreement for each species was defined by an overlap between error bars, and a logarithmic distance of disagreement was computed for each species in the absence of overlap.  Due to the large number of parameters in the gas-grain code, a full analysis of the error propagation would be very difficult to achieve, and has not so far been attempted.

Here we present a new means of comparison, whereby we assign a level of confidence in the agreement of each computed abundance (at a particular time) with the observed value, having designated it with a generic observational error. We construct a log-normal distribution about each observational value, and identify its defining standard deviation, $\sigma$, with an appropriate error factor on the observed value. In a similar way as one might determine the confidence level of a spectroscopic detection being distinct from the mean baseline, we conversely define our ``confidence", $\kappa_{i}$, that the calculated value, $X_{i}$, for species $i$ is {\em associated} with the observed value, $X_{obs,i}$, thus:
\begin{equation}
\kappa_{i} = \mbox{{\em erfc}} \left( \frac{| \log{\left(X_{i}\right)} - \log{\left(X_{obs,i}\right)} |}{\sqrt{2} \sigma} \right)
\end{equation}
\noindent where {\em erfc} is the complementary error function ({\em erfc} = 1 - {\em erf}); $\kappa_{i}$ ranges between zero and unity. The term ``confidence'' is contained within quotation marks to signify that we are not dealing with a rigorous statistical analysis. For our analysis, we define $\sigma = 1$, hence 1 standard deviation corresponds to one order of magnitude higher or lower than the observed value. Therefore, a calculated value which lies 1 order of magnitude from $X_{i,obs}$ has a confidence level $\kappa_{i} = 0.317$, 
whilst a value 2 orders of magnitude from $X_{obs,i}$ has a confidence level $\kappa_{i} = 0.046$. 
For each time in the model, we take a mean of the individual confidence levels of every species, allowing us to define the overall confidence in a match with observations for each model at each time.

\begin{figure}
\centering
\includegraphics[width=8.5cm]{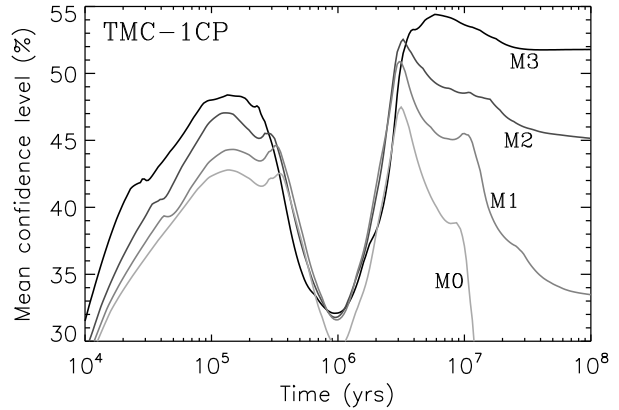}
\caption{Mean confidence level (\%) for fit with TMC-1CP observations.}
\label{fig99}
\end{figure}

Other means of comparing quantitatively with observations were considered; for example, taking the sum or mean average of $| \log{X_{i}} - \log{X_{obs,i}} |$ over all species, in a similar approach to that of \cite{wakelam06b}, to provide a parameter (the mean ``deviation'') which becomes smaller with a closer match. However, this method tends to skew results to some degree when there are ``large'' deviations from the observed values. For example, if in one model a particular species is 3 orders of magnitude from the observed value, and in another model, that species is 6 orders of magnitude away, the influence of this species on each mean will be large, even though in both cases the actual value is so far away as to be no longer credible. In such a case, we would suggest that the model itself is flawed (as regards that species) and that no amount of ``tuning'' can reasonably be expected to produce the right result (for the right reasons). But this would not necessarily make either model catastrophically wrong, in spite of what the comparison parameter might suggest.  The ``yes or no'' method employed by \cite{smith04a} naturally discards such outliers.  Our preferred method, as explained above, allows us to define, in a quasi-statistical manner,  the extent to which variations in calculated values are important when the values themselves are far from the target. Whilst our method is no more statistically rigorous than the alternatives adumbrated above,  it provides a convenient and self-consistent means of analysis that is easily automated, and which allows quantitative comparison between different models.

We now compare the results with detected abundances obtained for L134N and TMC-1CP (the so-called cyanopolyyne peak). Given in Tables \ref{tab3} and \ref{tab4}, these values are taken from a number of different studies, but correspond to those collated by \cite{wakelam06b}, for L134N, and \cite{smith04a}, for TMC-1CP.  To those species that have only an upper or lower limit, we assign a confidence level of unity if the calculated value is lower than the upper limit, or greater than the lower limit, respectively; they are otherwise treated in the same way.

\begin{table*}
\caption[]{Observational values of fractional abundances in L134N, and corresponding values at the ``best fit'' times for each model. }
\scriptsize
\label{tab3}
\begin{center}
\begin{tabular}{lccccc}
\hline
\hline
\noalign{\smallskip}
 Species & $N(i) / N($H$_{2})$ $^{1,2}$ & \multicolumn{4}{c}{$n(i) / n($H$_{2})$ $^{1,3}$} \\
\noalign{\smallskip}
 & Observed & M0 & M1 & M2 & M3 \\
\noalign{\smallskip}
 & (L134N) & $t=2.6$ Myr & $t=1.9$ Myr & $t=2.8$ Myr & $t=3.1$ Myr \\
\noalign{\smallskip}
\hline
\noalign{\smallskip}
C            & $\geq$1.0(-6) $^{4}$ &  6.2(-7) & {\bfseries  8.1(-8)} &  5.3(-7) &  2.0(-7) \\
\noalign{\smallskip}
CH           & 1.0(-8)  $^{5}$ &  6.8(-9) &  3.2(-9) &  7.7(-9) &  7.7(-9) \\
\noalign{\smallskip}
C$_2$H       & $\leq$5.0(-8)  $^{5}$ &  8.5(-9) &  5.5(-9) &  9.3(-9) &  1.1(-8) \\
\noalign{\smallskip}
C$_3$H       & 3.0(-10)  $^{5}$ & {\bfseries \em  6.8(-9)} &  1.5(-9) & {\bfseries \em  8.6(-9)} & {\bfseries \em  6.9(-9)} \\
\noalign{\smallskip}
C$_3$H$_2$   & 2.0(-9)  $^{5}$ & {\bfseries  1.0(-10)} & {\bfseries  3.6(-11)} &  2.2(-10) &  3.9(-10) \\
\noalign{\smallskip}
C$_3$H$_4$   & $\leq$1.2(-9)  $^{5}$ &  4.0(-11) &  2.4(-10) &  3.5(-10) &  1.4(-9) \\
\noalign{\smallskip}
C$_4$H       & 1.0(-9)  $^{5}$ &  3.9(-9) &  1.1(-9) &  4.7(-9) &  5.3(-9) \\
\noalign{\smallskip}
CN           & 8.2(-10)  $^{5}$ & {\bfseries \em  3.3(-8)} & {\bfseries \em  1.4(-8)} & {\bfseries \em  4.3(-8)} & {\bfseries \em  4.8(-8)} \\
\noalign{\smallskip}
HCN          & 1.2(-8)  $^{6}$ &  4.5(-8) &  2.0(-8) &  6.9(-8) &  8.3(-8) \\
\noalign{\smallskip}
HNC          & 4.7(-8) $^{6}$ &  4.0(-8) &  1.9(-8) &  6.3(-8) &  7.5(-8) \\
\noalign{\smallskip}
H$_2$CN$^+$  & $\leq$3.1(-9) $^{5}$ &  8.3(-10) &  3.6(-10) &  1.1(-9) &  1.2(-9) \\
\noalign{\smallskip}
CH$_2$CN     & $\leq$1.0(-9) $^{5}$ &  3.3(-10) &  1.0(-10) &  6.1(-10) &  8.8(-10) \\
\noalign{\smallskip}
CH$_3$CN     & $\leq$1.0(-9) $^{5}$ &  1.1(-10) &  2.3(-11) &  2.0(-10) &  2.4(-10) \\
\noalign{\smallskip}
C$_3$N       & $\leq$2.0(-10) $^{5}$ &  4.3(-10) &  1.0(-10) &  5.5(-10) &  5.5(-10) \\
\noalign{\smallskip}
C$_3$H$_3$N  & $\leq$1.0(-10) $^{5}$ &  1.6(-12) &  7.7(-13) &  5.7(-12) &  2.3(-11) \\
\noalign{\smallskip}
HC$_3$N      & 8.7(-10) $^{6}$ &  2.2(-10) &  1.1(-10) &  4.3(-10) &  8.5(-10) \\
\noalign{\smallskip}
HC$_5$N      & 1.0(-10) $^{5}$ &  2.2(-11) & {\bfseries  7.5(-12)} &  4.6(-11) &  1.2(-10) \\
\noalign{\smallskip}
HC$_7$N      & 2.0(-11) $^{5}$ & {\bfseries  5.9(-13)} & {\bfseries  1.4(-13)} & {\bfseries  1.6(-12)} &  5.1(-12) \\
\noalign{\smallskip}

CO           & 8.0(-5) $^{5}$ &  1.2(-5) &  2.2(-5) &  1.4(-5) &  2.3(-5) \\
\noalign{\smallskip}
HCO$^+$      & 1.0(-8) $^{6}$ &  2.1(-9) &  3.0(-9) &  2.3(-9) &  3.0(-9) \\
\noalign{\smallskip}
H$_2$CO      & 2.0(-8) $^{5}$ &  2.1(-8) &  1.3(-8) &  3.7(-8) &  7.8(-8) \\
\noalign{\smallskip}
CH$_3$OH     & 3.7(-9) $^{6}$ & {\bfseries  6.0(-13)} &  1.7(-9) &  4.1(-9) &  1.9(-8) \\
\noalign{\smallskip}
HCOOH        & 3.0(-10) $^{5}$ & {\bfseries  1.5(-11)} &  1.5(-10) &  3.0(-11) &  1.8(-10) \\
\noalign{\smallskip}
CH$_2$CO     & $\leq$7.0(-10) $^{5}$ &  2.9(-10) &  3.9(-10) &  3.2(-10) &  4.1(-10) \\
\noalign{\smallskip}
CH$_3$CHO    & 6.0(-10) $^{5}$ & {\bfseries  1.3(-13)} & {\bfseries  5.6(-13)} & {\bfseries  1.3(-12)} & {\bfseries  7.0(-12)} \\
\noalign{\smallskip}
C$_3$O       & $\leq$5.0(-11) $^{5}$ &  1.0(-11) &  7.2(-12) &  1.4(-11) &  2.1(-11) \\
\noalign{\smallskip}
H$_2$S       & 8.0(-10) $^{5}$ & {\bfseries  1.6(-11)} &  2.9(-9) &  3.1(-9) &  3.1(-9) \\
\noalign{\smallskip}
SO           & 3.1(-9) $^{6}$ & {\bfseries  6.9(-11)} &  3.2(-9) &  1.5(-9) &  4.4(-9) \\
\noalign{\smallskip}
SO$_2$       & $\leq$1.6(-9) $^{6}$ &  8.9(-13) &  7.6(-11) &  3.2(-11) &  3.3(-10) \\
\noalign{\smallskip}
CS           & 1.7(-9) $^{6}$ &  6.8(-10) &  6.1(-9) &  1.3(-8) & {\bfseries \em  1.9(-8)} \\
\noalign{\smallskip}
HCS$^+$      & 6.0(-11) $^{5}$ & {\bfseries  2.7(-12)} &  2.1(-11) &  5.0(-11) &  6.3(-11) \\
\noalign{\smallskip}
H$_2$CS      & 6.0(-10) $^{5}$ &  2.0(-10) &  1.1(-9) &  5.1(-9) & {\bfseries \em  8.4(-9)} \\
\noalign{\smallskip}
C$_2$S       & 6.0(-10) $^{5}$ & {\bfseries  2.5(-11)} &  1.5(-10) &  3.5(-10) &  2.6(-10) \\
\noalign{\smallskip}
C$_3$S       & $\leq$2.0(-10) $^{5}$ &  3.7(-12) &  1.4(-11) &  5.6(-11) &  4.6(-11) \\
\noalign{\smallskip}
OCS          & 2.0(-9) $^{5}$ & {\bfseries  4.0(-12)} & {\bfseries  6.2(-11)} & {\bfseries  7.5(-11)} & {\bfseries  1.4(-10)} \\
\noalign{\smallskip}
NH$_3$       & 9.1(-8) $^{5}$ &  1.7(-7) &  1.2(-7) &  2.3(-7) &  2.7(-7) \\
\noalign{\smallskip}
N$_2$H$^+$   & 6.8(-10) $^{6}$ &  2.3(-9) &  2.1(-9) &  2.5(-9) &  2.2(-9) \\
\noalign{\smallskip}
NO           & 6.0(-8) $^{5}$ &  4.4(-8) &  9.1(-8) &  5.9(-8) &  1.4(-7) \\
\noalign{\smallskip}
OH           & 7.5(-8) $^{5}$ &  3.6(-8) &  4.9(-8) &  4.9(-8) &  1.1(-7) \\
\noalign{\smallskip}
H$_2$O       & $\leq$3.0(-7) $^{7}$ &  4.5(-8) &  8.6(-8) &  9.1(-8) &  2.6(-7) \\
\noalign{\smallskip}
O$_2$        & $\leq$1.7(-7) $^{8}$ &  1.8(-7) & {\bfseries \em  2.6(-6)} &  2.0(-7) &  8.8(-7) \\
\noalign{\smallskip}
\hline
\noalign{\smallskip}
Matches$^{9}$ /41 && 29 & 33 & 36 & 35 \\
\noalign{\smallskip}
\hline
\hline
\noalign{\smallskip}
\multicolumn{6}{l}{$^{1}$ $a(b) = a \times 10^{b}$; $^{2}$ Observed values collated by \cite{wakelam06b}; $^{3}$ Boldface indicates a } \\
\multicolumn{6}{l}{theoretical value different by more than 1 order of magnitude from the observed value. Plain} \\
\multicolumn{6}{l}{boldface indicates too low a value; italic boldface indicates too high a value; $^{4}$ \cite{stark96a}; } \\
\multicolumn{6}{l}{$^{5}$ \cite{ohishi92a}; $^{6}$ \cite{dickens00a}; $^{7}$ \cite{snell00a}; $^{8}$ \cite{pagani03a}} \\
\multicolumn{6}{l}{$^{9}$ Agreement with observational value, to within 1 order of magnitude.} \\
\end{tabular}
\end{center}
\end{table*}

\begin{table*}
\caption[]{Observational values of fractional abundances in TMC-1CP, and corresponding values at the ``best fit'' times for each model. }
\scriptsize
\label{tab4}
\begin{center}
\begin{tabular}{lccccc}
\hline
\hline
\noalign{\smallskip}
 Species & $N(i) / N($H$_{2})$ $^{1,2}$ & \multicolumn{4}{c}{$n(i) / n($H$_{2})$ $^{1,3}$} \\
\noalign{\smallskip}
 & Observed & M0 & M1 & M2 & M3 \\
\noalign{\smallskip}
 & (TMC-1CP) & $t=3.2$ Myr & $t=3.1$ Myr & $t=3.3$ Myr & $t=6.0$ Myr \\
\noalign{\smallskip}
\hline
\noalign{\smallskip}
CH      &    2(-8)   &      8.5(-9)         &      8.7(-9)         &      9.0(-9)         &      8.7(-9)      \\
\noalign{\smallskip}
C$_2$      &    5(-8)   &      4.4(-8)         &      3.8(-8)         &      5.3(-8)         &      2.4(-7)      \\
\noalign{\smallskip}
C$_2$H     &    2(-8)   &      8.1(-9)         &      8.2(-9)         &      8.4(-9)         &      7.3(-9)      \\
\noalign{\smallskip}
C$_3$H     &    1(-8)   &      1.2(-8)         &      1.2(-8)         &      1.5(-8)         &      3.0(-8)      \\
\noalign{\smallskip}
C$_3$H$_2$    &    1(-8)   & {\bfseries 1.9(-10)} & {\bfseries 2.3(-10)} & {\bfseries 4.0(-10)} &      2.6(-9)      \\
\noalign{\smallskip}
C$_3$H$_4$    &    6(-9)   & {\bfseries 7.4(-11)} & {\bfseries 1.7(-10)} & {\bfseries 4.2(-10)} &      6.4(-9)      \\
\noalign{\smallskip}
C$_4$H     &    9(-8)   & {\bfseries 5.9(-9)}  & {\bfseries 5.6(-9)}  & {\bfseries 7.7(-9)}  &      4.9(-8)      \\
\noalign{\smallskip}
C$_4$H$_2$    &    1(-9)   &      2.3(-9)         &      2.2(-9)         &      3.2(-9)         & {\bfseries \em 2.3(-8)}   \\
\noalign{\smallskip}
C$_5$H     &    6(-10)  &      7.5(-10)        &      7.7(-10)        &      1.1(-9)         &      4.9(-9)      \\
\noalign{\smallskip}
CH$_3$C$_4$H  &    4(-10)  &      4.9(-11)        &      5.5(-11)        &      9.8(-11)        &      1.7(-9)      \\
\noalign{\smallskip}
C$_6$H     &    2(-10)  &      2.0(-10)        &      2.0(-10)        &      3.3(-10)        & {\bfseries \em 3.2(-9)}   \\
\noalign{\smallskip}
C$_6$H$_2$    &    5(-11)  &      5.4(-11)        &      6.0(-11)        &      1.1(-10)        & {\bfseries \em 1.7(-9)}   \\
\noalign{\smallskip}
CN      &    5(-9)   &      5.0(-8)         & {\bfseries \em 5.1(-8)} & {\bfseries \em 6.4(-8)} & {\bfseries \em 1.2(-7)}   \\
\noalign{\smallskip}
HCN     &    2(-8)   &      7.8(-8)         &      8.5(-8)         &      1.1(-7)         & {\bfseries \em 2.7(-7)}   \\
\noalign{\smallskip}
HNC     &    2(-8)   &      7.1(-8)         &      7.6(-8)         &      1.0(-7)         & {\bfseries \em 2.5(-7)}   \\
\noalign{\smallskip}
H$_2$CN$^+$   &    2(-9)   &      1.5(-9)         &      1.5(-9)         &      1.8(-9)         &      3.2(-9)      \\
\noalign{\smallskip}
CH$_2$CN   &    5(-9)   &      7.0(-10)        &      7.9(-10)        &      1.1(-9)         &      3.6(-9)      \\
\noalign{\smallskip}
CH$_3$CN   &    6(-10)  &      2.6(-10)        &      2.9(-10)        &      4.1(-10)        &      1.1(-9)      \\
\noalign{\smallskip}
C$_3$N     &    6(-10)  &      7.0(-10)        &      7.2(-10)        &      9.5(-10)        &      3.8(-9)      \\
\noalign{\smallskip}
HNC$_3$    &    6(-11)  &      3.9(-11)        &      4.1(-11)        &      5.8(-11)        &      2.8(-10)      \\
\noalign{\smallskip}
HC$_2$NC   &    5(-10)  & {\bfseries 3.2(-11)} & {\bfseries 3.5(-11)} &      5.3(-11)        &      3.6(-10)      \\
\noalign{\smallskip}
C$_3$H$_2$N$^+$  &    1(-10)  & {\bfseries 8.7(-12)} & {\bfseries 8.7(-12)} &      1.2(-11)        &      5.9(-11)      \\
\noalign{\smallskip}
C$_3$H$_3$N   &    4(-9)   & {\bfseries 4.8(-12)} & {\bfseries 6.1(-12)} & {\bfseries 1.4(-11)} & {\bfseries 3.0(-10)}   \\
\noalign{\smallskip}
CH$_3$C$_3$N  &    8(-11)  & {\bfseries 8.4(-13)} & {\bfseries 9.8(-13)} & {\bfseries 1.6(-12)} & {\bfseries 6.8(-12)}   \\
\noalign{\smallskip}
HC$_3$N    &    2(-8)   & {\bfseries 4.7(-10)} & {\bfseries 5.2(-10)} & {\bfseries 8.5(-10)} &      6.2(-9)      \\
\noalign{\smallskip}
HC$_5$N    &    4(-9)   & {\bfseries 4.8(-11)} & {\bfseries 5.1(-11)} & {\bfseries 9.8(-11)} &      2.2(-9)      \\
\noalign{\smallskip}
HC$_7$N    &    1(-9)   & {\bfseries 1.7(-12)} & {\bfseries 1.9(-12)} & {\bfseries 4.3(-12)} &      1.8(-10)      \\
\noalign{\smallskip}
HC$_9$N    &    5(-10)  & {\bfseries 1.6(-13)} & {\bfseries 1.7(-13)} & {\bfseries 4.2(-13)} & {\bfseries 4.0(-11)}   \\
\noalign{\smallskip}
CO      &    8(-5)   & {\bfseries 7.6(-6)}  &      8.7(-6)         &      1.0(-5)         &      9.5(-6)      \\
\noalign{\smallskip}
HCO$^+$    &    8(-9)   &      1.5(-9)         &      1.6(-9)         &      1.7(-9)         &      1.0(-9)      \\
\noalign{\smallskip}
H$_2$CO    &    5(-8)   &      2.5(-8)         &      2.9(-8)         &      3.8(-8)         &      6.2(-8)      \\
\noalign{\smallskip}
CH$_3$OH   &    3(-9)   & {\bfseries 1.0(-12)} &      1.1(-9)         &      3.7(-9)         &      1.1(-8)      \\
\noalign{\smallskip}
HCOOH   & $<$2(-10)  &      5.8(-12)        &      9.6(-12)        &      1.8(-11)        &      8.5(-11)      \\
\noalign{\smallskip}
CH$_2$CO   &    6(-10)  &      1.8(-10)        &      2.1(-10)        &      2.3(-10)        &      2.3(-10)      \\
\noalign{\smallskip}
CH$_3$CHO  &    6(-10)  & {\bfseries 1.5(-13)} & {\bfseries 4.4(-13)} & {\bfseries 1.3(-12)} & {\bfseries 1.7(-11)}   \\
\noalign{\smallskip}
C$_2$O     &    6(-11)  & {\bfseries 4.6(-12)} & {\bfseries 5.5(-12)} &      7.7(-12)        &      8.9(-12)      \\
\noalign{\smallskip}
C$_3$O     &    1(-10)  & {\bfseries 8.7(-12)} &      1.0(-11)        &      1.6(-11)        &      1.3(-10)      \\
\noalign{\smallskip}
H$_2$S     & $<$5(-10)  &      1.2(-11)        &      3.0(-9)         &      3.0(-9)         &      7.8(-10)      \\
\noalign{\smallskip}
SO      &    2(-9)   & {\bfseries 3.0(-11)} &      8.1(-10)        &      1.2(-9)         &      1.4(-9)      \\
\noalign{\smallskip}
SO$_2$     & $<$1(-9)   &      2.8(-13)        &      9.9(-12)        &      2.2(-11)        &      1.9(-10)      \\
\noalign{\smallskip}
CS      &    4(-9)   &      6.9(-10)        &      1.5(-8)         &      2.0(-8)         &      2.0(-8)      \\
\noalign{\smallskip}
HCS$^+$    &    4(-10)  & {\bfseries 3.0(-12)} &      6.1(-11)        &      7.6(-11)        &      6.2(-11)      \\
\noalign{\smallskip}
H$_2$CS    &    7(-10)  &      2.6(-10)        &      5.9(-9)         & {\bfseries \em 8.0(-9)}   & {\bfseries \em 8.6(-9)}   \\
\noalign{\smallskip}
C$_2$S     &    8(-9)   & {\bfseries 2.3(-11)} & {\bfseries 4.6(-10)} & {\bfseries 5.3(-10)} & {\bfseries 3.9(-10)}   \\
\noalign{\smallskip}
C$_3$S     &    1(-9)   & {\bfseries 3.7(-12)} & {\bfseries 7.4(-11)} & {\bfseries 9.1(-11)} & {\bfseries 9.2(-11)}   \\
\noalign{\smallskip}
OCS     &    2(-9)   & {\bfseries 2.1(-12)} & {\bfseries 5.1(-11)} & {\bfseries 6.1(-11)} & {\bfseries 2.8(-11)}   \\
\noalign{\smallskip}
NH$_3$     &    2(-8)   & {\bfseries \em 2.3(-7)} & {\bfseries \em 2.4(-7)} & {\bfseries \em 2.7(-7)} & {\bfseries \em 3.6(-7)}   \\
\noalign{\smallskip}
N$_2$H$^+$    &    4(-10)  &      2.3(-9)         &      2.4(-9)         &      2.4(-9)         &      1.4(-9)      \\
\noalign{\smallskip}
NO      & $<$3(-8)   &      2.8(-8)         &      3.4(-8)         &      4.1(-8)         &      9.4(-8)      \\
\noalign{\smallskip}
OH      &    2(-7)   &      2.6(-8)         &      3.1(-8)         &      4.3(-8)         &      1.6(-7)      \\
\noalign{\smallskip}
H$_2$O     & $\leq$7.0(-8) &   3.4(-8)         &      5.1(-8)         &      9.0(-8)         &      5.0(-7)      \\
\noalign{\smallskip}
O$_2$      & $\leq$7.7(-8) &   4.0(-8)         &      5.5(-8)         &      6.4(-8)         &      8.4(-8)      \\
\noalign{\smallskip}
\hline
\noalign{\smallskip}
Matches$^{4}$ /52 && 30 & 34 & 36 & 37 \\
\noalign{\smallskip}
\hline
\hline
\noalign{\smallskip}
\multicolumn{6}{l}{$^{1}$ $a(b) = a \times 10^{b}$; $^{2}$ Observed values collated by \cite{smith04a}; $^{3}$ Boldface indicates a } \\
\multicolumn{6}{l}{theoretical value different by more than 1 order of magnitude from the observed value. Plain} \\
\multicolumn{6}{l}{boldface indicates too low a value; italic boldface indicates too high a value.} \\
\multicolumn{6}{l}{$^{4}$ Agreement with observational value, to within 1 order of magnitude.} \\
\end{tabular}
\end{center}
\end{table*}

\subsection{L134N}

Figure \ref{fig98} shows the level of confidence in the match between each model and the 41 observational values for L134N, for times $10^{4}$ -- $10^{8}$ yr. Clearly the models M1 -- M3 are an improvement over the values obtained without the new mechanism. The best match at any time is obtained for model M2 ($a=0.03$), at a time of $\sim$3 Myr; however, each of models M1 -- M3 reaches its peak at around this time, and the differences in confidence are very small. The model M2 peak is also the widest, spanning around 1 Myr with a $\sim$69\% 
confidence level. Each of the models shows a confidence of $\sim$$65$ \% over a wider time interval of 
 $t = 10^{5}$ -- $3 \times 10^{6}$ yr. This corresponds to less than a factor of 3 deviation, on average, between modelled and observed abundances.
Model M3 in particular has another strong peak, at $\sim$$10^{5}$ yr, but M2 also shows a local peak at around this time. An age for L134N of anywhere between 0.1 and 3 Myr may therefore be plausible according to models M1 -- M3; methanol is still adequately produced at all such times. Whilst M3 also has a very high agreement at 10$^{8}$ yr, on dynamical grounds we should consider a match at this time to be highly dubious.

Clearly, whilst the introduction of the new mechanism much improves the agreement with observed abundances according to our method of analysis, variation of the parameter $a$ in the $0.01 - 0.1$ range does not strongly alter it, although higher values make an earlier match more of a possibility. The mean confidence level appears therefore to give us only limited scope for the constraint of the $a$-parameter.  Comparing the mean confidence level approach with a method that utilizes the mean of the log of the deviation, we find the latter to yield similar times of best agreement, but to suggest that M3 (as opposed to M2) achieves the best match with observations.  The difference is marginal, though, and serves to demonstrate the difficulty in constraining the value of $a$. 

Table \ref{tab3} shows the abundances of molecules observed in L134N and the calculated values for each model at its own peak at $\sim$2 Myr. Values which vary from the detected value by more than one order of magnitude are printed in boldface. Values which exceed the detected level by this much are also set in italics. 

Aside from SO, H$_2$S and O$_2$, all of the highlighted discrepancies occur in the abundances of carbon-bearing species. Indeed, most of these occur for species containing more than 1 carbon atom. The larger cyanopolyynes are not well matched at the peak times, typically falling quite short of observed values.  At later times (close to 10 Myr), however, these abundances are fairly well matched with observations. CN is badly overproduced in all models. Hydrocarbon abundances are typically raised for higher $a$-values, although, except for C$_3$H, they are usually still a fair match. The best-matched time for larger $a$ also tends to under-represent atomic C, though not by more than 1 order of magnitude. 
For the most part, those species with observed abundances not well-reproduced by model M0 are either much improved upon with all of models M1 -- M3, or barely improved at all -- much as Fig. \ref{fig98} represents. Very few species are actually worsened in their agreement by the new mechanism. The best match to observations with the order-of-magnitude criterion is still model M2, with an age of around 2 - 3 Myr, the same as determined by the mean confidence level. However, earlier matches, to as short an age as 10$^5$ yr, are plausible, and methanol is still adequately produced at such times.

 It is worth comparing the success of the gas-grain code with that of our most recent gas-phase code.  \cite{smith06a} achieved 73\% 
of calculated abundances falling within 1 order of magnitude of the observed value, for L134N, at time 10$^5$ yr. Our best model achieves 88\% 
(36/41 species), at $\sim$2 Myr. Clearly, our new gas-grain models improve agreement with observations over a purely gas-phase analysis, and this is true over the entire range of $\sim$$10^5$ -- few $\times 10^6$ yr. However, in the case where $a = 0$, model M0, a comparable level of agreement with the gas-phase treatment (70\%) is achieved.

\subsection{TMC-1CP}

Figure \ref{fig99} shows the mean confidence levels for each model in comparison to 52 species detected in TMC-1CP. The low mean confidence obtained at all times is apparent, and corresponds to a factor of 5 -- 10 deviation on average, significantly higher than for L134N. Nevertheless, we can clearly see that greater $a$-values produce a better match with the TMC-1 values. Also, there are two peaks, but the later-time peak is much stronger, especially for greater values of $a$. This stands in contrast to the commonly-held view of TMC-1 as a relatively young object \cite[]{hartquist01a}.

Table \ref{tab4} shows the observed and computed best-time abundances for each molecule in the observational data set. We can see immediately that most molecules that fail to be reproduced fail to some degree in almost every model. As for L134N, the majority of these are carbon-bearing species. Some of these species achieve an acceptable match with observations using model M3 ($a=0.1$), however, other carbon-bearing species then attain levels which are too high. For other species, model M3 merely improves upon the other models, without getting reliably close to the observed value. In the case of models M0 -- M2, we might also argue that the fact that so many values fall below the observed values, yet so few lie above, makes it unlikely that the discrepancies are purely a statistical effect; however, model M3 is well-balanced in this respect. Test runs using values yet higher than $a=0.1$ produce an ever better match, however, such values would be very difficult to justify on physical grounds, and the fact that their agreement peaks at even later times makes the match less plausible. The implication is not therefore that M3 is a significantly better model of TMC-1CP, but that each model has failed, model M3 having failed only to a lesser degree. Dismissing the late-time peak on chemical and dynamical grounds would therefore suggest an optimum age of around 1 -- 3 $\times$ 10$^5$ yr, in keeping with typical estimates.

One possible reason for the failure of our models is a lack of gas-phase carbon; \cite{smith04a} tested increases in the C:O ratio for the gas-phase model, and found improvements in the agreement with TMC-1CP, which by all accounts appears to have an unusual chemistry. New estimates of carbon and oxygen abundances \cite[]{sofia94,meyer98} are both somewhat higher than those used here, but are unlikely to alter the C:O ratio by a very great degree. As found by \cite{gwenlan00a}, a much higher cosmic-ray ionisation rate may also contribute to the unusual chemistry of TMC-1. In any case, it is likely that we are lacking a crucial element in any attempt to model TMC-1CP.

\subsection{Grain mantle composition}

\begin{table}
\caption{Observed ice composition towards Elias 16, expressed as a percentage of H$_2$O abundance. $^1$}
\label{tab5}
\begin{center}
\begin{tabular}{lccc}
\hline
\hline
Species  & Elias 16 & \multicolumn{2}{c}{Model M2}  \\
         &          & $2\times10^{5}$ yr & $2.8\times10^{6}$ yr \\
\hline
H$_2$O $^{2,3}$ & 1.28(-4) & 3.47(-5) & 1.50(-4) \\
CO $^3$ & 26 & 7.4 & 0.3 \\
CO$_2$ $^4$ & 20 & 4(-2) & 0.3 \\
CH$_4$ $^5$ & - & 20.7 & 28.5 \\
CH$_3$OH $^5$ & $<3$ & 6.4 & 10.5 \\
H$_2$CO $^5$ & - & 5.7 & 1.0 \\
H$_2$O$_2$ $^5$ & $<5$ & 3(-4) & 5(-5) \\
OCS $^5$ &  $<0.2$ & 9(-13) & 6(-14) \\
NH$_3$ $^5$ & $<9$ & 12.6 & 7.3 \\
\hline
\end{tabular}
\end{center}
$^{1}$ $a(b) = a \times 10^{b}$; 
$^{2}$ H$_2$O fractional abundance; 
$^{3}$ ref. \cite{nummelin01a}; 
$^{4}$ ref. \cite{bergin05a}; 
$^{5}$ data collated by \cite{gibb00a}
\end{table}

The determination of ice composition in dark clouds is dependent on the absorption of infrared radiation from background field stars. Unfortunately, no data exist for lines of sight directly towards L134N or TMC-1CP. However, a number of studies have concentrated on field stars behind the Taurus complex of dark clouds. Such data may give some indication of the likely grain mantle composition in the objects we have modelled. We compare our results with the ice composition determined towards background star Elias 16, which lies in the general vicinity of TMC-1CP. Visual extinction along this line of sight is determined to be around $A_{V} = 24$ \cite[]{whittet07a}.

Table \ref{tab5} shows the observed composition of icy mantles as a percentage of detected water ice, towards Elias 16. As the modelled grain mantle abundances are generally not very sensitive to the new desorption mechanism, we present results for model M2, chosen as representative of all models, at time $2.8 \times 10^{6}$ yr, corresponding to the late-time peak in the agreement of gas phase species with observations of L134N, and at time $2 \times 10^{5}$ yr, corresponding approximately with the early-time peak in agreement with both L134N and TMC-1CP observations (see figures \ref{fig98} and \ref{fig99}).

Whilst water abundance is fairly well-reproduced, there is clearly no strong agreement at either time-value for the strongest carbon-bearing constituents. CO and CO$_2$ are under-represented in the model, with CH$_4$ comprising much of this carbon. Methanol is also somewhat more abundant in our models than the upper limit of the observations suggest; however, a moderate reduction should have no major effect on the gas-phase abundance through the new mechanism. Formaldehyde is not detected, whereas quantities on the order of a percent are formed in our models. Ammonia is in acceptable agreement with observations, and H$_2$O$_2$ and OCS fall comfortably below their upper limits.

The fact that the models do not well reproduce the main features of the observations is perhaps not surprising; previous models using the canonical dark cloud physical conditions have not typically been very successful in this respect. \cite{ruffle01b} attempted to rectify this, running a grid of models at various temperatures and densities and adopting different grain surface binding energies and diffusion barriers. They found that greater temperatures and/or greater densities could account for the particular surface composition seen in Elias 16, especially in reproducing CO$_2$. We might therefore suggest that the observed ice composition is representative of a marginally higher temperature, say 15K. In our model this would lead to lower residence times for atomic hydrogen, resulting in less efficient hydrogenation. Our model adopts a set of binding energies and diffusion barriers representative of amorphous water ice; our atomic hydrogen diffusion barrier is a little higher than the lowest used by Ruffle \& Herbst (225 K cf. 200 K), and our binding energy is rather higher than their lowest (450 K cf. 373 K). Our model also uses their low activation energy (80 K) for the CO + O $\rightarrow$ CO$_2$ surface reaction. A lower hydrogenation rate (due to a higher temperature) would result in less CH$_4$ formation, and would allow CO$_2$ formation to compete with hydrogenation, as in the Ruffle \& Herbst models. However, we might then argue that the Elias 16 observations are not a fair comparison, as temperatures may well be different from those of the regions we are modelling. 

The apparent disagreement between the model and observations should also be put into the context of our comparison with gas-phase abundances. Assuming a factor of 10 criterion for a successful match with observations, our ice results would appear much more acceptable, especially for earlier times of agreement.

Nevertheless, the strong surface CH$_4$ abundance we find is not representative of observations of various other objects, typically reaching no more than a few percent of water ice \cite[]{gibb00a}. As suggested by \cite{bergin05a}, surface CH$_4$ formation may be suppressed by a degree of gas-phase CO formation at earlier times, locking up carbon before significant grain mantles form. This would also favour greater CO$_2$ formation, occurring in tandem with the hydrogenation of oxygen to H$_2$O. Such a scenario agrees with the best fits to the ice observations of \cite{bergin05a}, which indicate that the majority of the CO$_2$ resides in the deeper, polar layers. So the initial stages of evolution of the gas may be crucial to the ultimate composition of the ice, and indeed a static model such as this one may not be entirely adequate.

\section{Conclusions}

The new desorptive mechanism, based on RRK theory \citep{holbrook96a}, allows the chemical energy released from grain-surface addition reactions to break the molecule-surface bond of the product, with a small probability on the order of 1\% 
governed by a parameter $a$. Under dark cloud conditions, each of the elements C, O, N and S (plus CO) and their various hydrides may participate in fast grain-surface hydrogen-addition reactions, which may therefore result in desorption {\em via} the new mechanism. 

We delineate two main effects of the mechanism; an early-time and a late-time effect. At earlier times ($t < \sim$$2$ Myr), the appearances of gas-phase abundance profiles are in general not qualitatively altered, although in the case of a few select species, in particular, hydrogenated species such as methanol, the quantities are strongly enhanced. At late times, when depletion would otherwise be strong, the new mechanism acts to maintain gas-phase chemistry, by re-injecting the various hydrogenated species -- species that may achieve very large abundances on the grains. The re-injection occurs {\em via} photodissociation of stable surface species into radicals followed by addition reactions, which lead to desorption {\em via} the new mechanism. This process feeds the formation of other species that were well-represented in the gas-phase up until the onset of strong depletion. Because of this effect, full-scale depletion is not attained, and a steady state is reached.
It is important to note that in the case of CO, the strongest effect on both its gas-phase and grain-surface abundances, at late times, is not its own re-injection, but its gas-phase formation from surface hydrocarbons. 
Indeed, one of the most striking aspects of the late-time effect is the high abundance attained by grain-surface hydrocarbons. Our treatment of carbon chains in the gas-grain code, however,  is certainly not comprehensive. Therefore, the abundance of these species may be overestimated, although we should expect some organic residue. However, the absence of observational evidence in dark clouds of solid hydrocarbons more complex than CH$_4$ may suggest that the long timescales needed to acquire a significant hydrocarbon content in the grain mantles are not dynamically realistic. This would indicate that dark clouds, or the physical structures within them, are inherently short-lived, on the order of a few million years or less.

Our new models show a good level of agreement with observed abundances for L134N. All of the models M1 -- M3 show improved agreement over M0, the model without the new mechanism. It is hard to constrain the value of the $a$-parameter by comparison with the observations of L134N; the net variation in abundances between models M1 -- M3 is not great. On the basis of methanol, and the general level of agreement with L134N, we would suggest a value of around $a=0.03$ as the optimum, and probably the maximum. (Such a value would suggest an optimum age for L134N of $\sim$3 Myr, although we find that a much lesser age may also be plausible in this regime, perhaps as short as $\sim$10$^5$ yr.) However, the most credible determination of $a$ must ultimately be an experimental one. The simulations of \cite{kroes06a} are encouraging in that they exhibit an effect of a similar strength for H$_2$O. But due to the small size of the expected desorption fraction ($\sim1$\% or less), 
the mere detection of the effect may be beyond experimental means at this time.

Contrary to the case of L134N, observations of TMC-1CP are not well-matched by any of our models; the increasing agreement for continually higher $a$ indicates that our models are not capable of producing a good match at any time, probably because we lack some crucial information.  A model that includes dynamics/structure in the cloud may be required.  Rather than a quiescent cloud, the TMC-1 ridge may be better represented by a superposition of clumps in varying stages of evolution \citep{garrod05a,garrod06a,peng98a}. A possible explanation for the unique chemistry of TMC-1 is the explosive desorption of grain mantles caused by MHD activity \cite[]{markwick00a}.

Our models do not show strong agreement with the observed ice compositions determined towards Elias 16. In particular, our models produce significant amounts of CH$_4$, and much lower CO and CO$_2$ abundances. We suggest, after \cite{ruffle01b}, that a marginally greater temperature may better reproduce the measured ice composition; such may not be representative of the regions whose gas-phase abundances we model, however. Alternatively, greater gas-phase production of CO prior to significant mantle formation may also improve agreement \cite[]{bergin05a}.

Qiang, Cuppen \& Herbst (in prep.) have recently studied cold cores with a new gas-grain approach, in which the grain chemistry is treated by a stochastic method, which is able to distinguish contributions from individual monolayers. With this approach, the observed abundances of CO$_2$ and methanol ices in Elias 16 are fit reasonably at a time of 10$^{5}$ yr whether one starts from flat or rough olivine surfaces.  As in earlier work, a temperature of 15 K results in improved agreement.

Importantly, the inclusion of the new desorptive mechanism, over the range of $a$-values, does not drastically alter the observable chemistry of our models of quiescent dark cloud regions, but generally improves their agreement with observations, while suggesting somewhat closer agreement at longer timescales. Whilst comparison with dark cloud abundances has been largely fruitless in strongly constraining the $a$-parameter, a model of pre-stellar cores may produce more success. These cold, dense regions are typefied by their strong CO depletion \cite[e.g.][]{redman02a}. Therefore, whilst for our quiescent dark cloud models the time at which depletion strongly takes hold is dynamically very late, pre-stellar cores must by definition reach such a state within a dynamically achieveable period. We should compare chiefly against CO, due to its strong and well-documented depletion levels. A model that includes deuterium chemistry would also be instructive due to the strong dependence of deuterium fractionation on the level of CO depletion. The degree to which the new mechanism may hold off depletion will be of interest in determining the ages of such objects.

\begin{acknowledgements}

The authors thank the National Science Foundation for partial support of this work. We thank the referee for helpful comments.

\end{acknowledgements}

\bibliographystyle{aa}

\bibliography{robbib_2006c-v3EH}

\end{document}